\documentclass[12pt,thmsa,a4paper]{article}
\usepackage{epsfig}

\begin{document}

\hfill FTUV/01-0410

\hfill IFIC/01-21

\hfill MZ-TH/01-12

\hfill April 2001

\bigskip \bigskip

\begin{center}
{\LARGE QCD Duality and the Mass of the Charm Quark\footnote{{\LARGE 
{\footnotesize Supported by CICYT under contract AEN99/0692, EC-RTN under
contract HPRN-CT2000/130, Generalitat Valenciana under contract GV98-1-80
and partnership Mainz-Valencia Universities}}}}\vspace{2cm}

{\large J. Pe\~{n}arrocha}$^{a}${\large \ and K. Schilcher}$^{b}\bigskip $

$^{a}${\normalsize Departamento de F\'{i}sica Te\'{o}rica-IFIC, Universitat
de Valencia}

{\normalsize E-46100 Burjassot-Valencia, Spain}

$^{b}${\normalsize Institut f\"{u}r Physik,
Johannes-Gutenberg-Universit\"{a}t}

{\normalsize D-55099 Mainz, Germany\vspace{2cm}}

\textbf{Abstract\medskip }
\end{center}

\begin{quote}
The mass of the charm quark is analyzed in the context of QCD finite energy
sum rules using recent BESII $e^{+}e^{-}$ annihilation data and a large
momentum expansion of the QCD correlator which incorporates terms to order $%
\alpha _{s}^{2}(m_{c}^{2}/q^{2})^{6}$. Using various versions of duality, we
obtain the consistent result $m_{c}(m_{c})=(1.37\pm 0.09)GeV$. Our result is
quite independent of the ones based on the inverse moment analysis.

\bigskip
\end{quote}

{\LARGE \newpage }

\section{Introduction}

Recently the BESII collaboration has presented new data on the total $%
e^{+}e^{-}$ annihilation cross section above the charm threshold\ \cite
{bessII}. Data in this energy region are particularly relevant for the
extraction of the charm quark mass, one of the fundamental parameters of
QCD. \ The charm quark can be determined by comparing suitable positive
moments of these data with the corresponding moments of QCD perturbation
theory. This direct quark-hadron duality approach was originally applied to
the charm region in Ref. \cite{NdeR}. A reanalysis along these lines seems
indicated in view of the fact that apart from new data, enormous progress
has been made in the theoretical calculation of the relevant QCD correlator
in the region $q^{2}\gg m_{c}^{2}$. The correlator is now known to $O(\alpha
_{s}^{2})$ and $O(m^{12}/q^{12})$ \cite{Chet1}, so that the question of
convergence can be meaningfully discussed. There exist also a result to $%
O(\alpha _{s}^{3})$ for the quartic mass correction \cite{Chet2} which
we will not consider for reasons of consistency. We believe that in the
case of the charm quark mass the direct duality approach employed by us
is less prone to theoretical uncertainties as the more popular one
based on inverse moments
\cite{Jamin},\cite{Naris},\cite{NarisPL},\cite{Domin},\cite{pineda}.  It
should be pointed out, that we use as phenomenological input only the
new BESII data. This is because older data in this region are plagued by
unknown systematical errors and appear to be mutually inconsistent. An
alternative would have been to adjust the normalizations of the various
data sets so that they agree for large $q^{2}$ with QCD \cite{Kuhn}.

We will manipulate our data rather on the basis of QCD duality. Using
suitable linear combinations of moments the emphasis of the hadronic
integral can be shifted at will to experimental regions where the data
errors are small. This technique, which was originally proposed in
\cite{ACD}, allows in many cases more accurate prediction, and supplies
in addition beautiful consistency checks. This will be explained in the
following.

\section{Cauchy sum rule}

QCD duality means that the theoretical and phenomenological information is
being related by means of Cauchy sum rule 
\begin{equation}
\int_{s_{0}}^{R}\frac{1}{\pi }{\rm Im}\Pi (s)p(s)ds=-\frac{1}{2\pi i}%
\oint_{\left| s\right| =R}\Pi _{QCD}(s)p(s)ds  \label{srule}
\end{equation}
where ${\rm Im}\Pi (s)$ is defined in terms of the total $e^{+}e^{-}$
annihilation cross-section by 
\begin{equation}
R(e^{+}e^{-}\rightarrow hadrons)=12\pi \sum_{flavors}Q_{f}^{2}{\rm Im}\Pi
(s)  \label{Rhad}
\end{equation}
The experimental charm physical threshold in Eq.(\ref{srule}) is taken from
the $J/\Psi $ resonance mass 
\begin{equation}
s_{0}=\left( 3.097Gev\right) ^{2}  \label{s0}
\end{equation}
We have included in the sum rule a polynomial weight function $p(s)$%
\begin{equation}
p(s)=\sum a_{n}s^{n}  \label{pol}
\end{equation}
which makes the sum rule a linear combination of moments of Cauchy sum
rules. The polynomial may be chosen in a suitable way to enhance or remove
part of the phenomenological input in the calculation.

\section{QCD integral}

The two-point function $\Pi _{QCD}(s)$ is known to $O(\alpha _{s}^{2})$ as a
series expansion in powers of $m^{2}/s$ up the sixth power,

\begin{equation}
\Pi _{QCD}(s)=\sum_{i=0}^{6}\sum_{j=0}^{3}A_{ij}(m,\mu )\left( \frac{m^{2}}{s%
}\right) ^{i}\left( \ln \frac{-s}{\mu ^{2}}\right) ^{j}  \label{Iqcd}
\end{equation}
where the coefficients $A_{ij}(m,\mu )$ may contain powers of mass
logarithms $\ln (m^{2}/\mu ^{2})$. The convergence of the series is not
seriously affected by the mass logarithms provided 
\begin{equation}
\ln \frac{\mu ^{2}}{m^{2}}\ll \ln \frac{\mu ^{2}}{\Lambda _{QCD}^{2}}\,.
\label{conv}
\end{equation}
This is always the case for the scales of $\mu $ ($\sim 5GeV$) we use.

At tree level, the first few terms in the expansion of $\Pi _{QCD}(s)$ are
given by

\[
\Pi _{QCD}(s)=\frac{3}{16\pi ^{2}}\left[ \frac{20}{9}-\frac{4}{3}\ln \frac{-s%
}{\mu ^{2}}+8\frac{m^{2}}{s}+\left( \frac{m^{2}}{s}\right) ^{2}\left( 4-8\ln 
\frac{m^{2}}{\mu ^{2}}+8\ln \frac{-s}{\mu ^{2}}\right) +..\right] 
\]
We use the strong coupling constant $\alpha _{s}^{2}$ and running
$\overline{MS}$ mass renormalized at the scale $\mu$. The lengthy full
expression to $O(\alpha _{s}^{2})$ and $O(m^{12}/q^{12})$ may be found
in Ref.\cite{Chet1}.

There is also a small non-perturbative contribution arising from the gluon
condensate, which is known to $O(\alpha _{s})$ \cite{Gcond}. The first few
terms are 
\[
\Pi _{np}(s)=\left\langle \frac{\alpha _{s}}{\pi }GG\right\rangle \left[ -%
\frac{1}{12}\left( \frac{m^{2}}{s}\right) ^{2}-\frac{1}{3}\left( \frac{m^{2}%
}{s}\right) ^{3}-\left( \frac{m^{2}}{s}\right) ^{4}\left( -\frac{7}{3}-\ln 
\frac{m^{2}}{\mu ^{2}}+\ln \frac{-s}{\mu ^{2}}\right) +..\right] 
\]

The Cauchy-Integral in Eq.(\ref{srule}) needs the evaluation of 
\begin{equation}
J(k,j)=\frac{1}{2\pi i}\oint_{\left| s\right| =R}s^{k}\left( \ln \frac{-s}{%
\mu ^{2}}\right) ^{j}ds  \label{Icau}
\end{equation}
for $j=0,1,2,3$ and $k=-6,-5,-4,...$These integrals can be evaluated
analytically with the result

\begin{eqnarray*}
J(k &\neq &-1,j)=\frac{1}{2\pi i}\oint_{\left| s\right| =R}s^{k}\left( \ln 
\frac{-s}{\mu ^{2}}\right) ^{j}ds=\frac{R^{n+1}}{n+1}\times \\
&&\left[ j\ln ^{j-1}\frac{R}{\mu ^{2}}-\frac{j(j-1)}{n+1}\ln ^{j-2}\frac{R}{%
\mu ^{2}}-\left( \frac{\pi ^{2}}{6}-\frac{1}{(n+1)^{2}}\right)
j(j-1)(j-2)\ln ^{j-3}\frac{R}{\mu ^{2}}\right]
\end{eqnarray*}
and 
\begin{eqnarray*}
J(k &=&-1,j)=\frac{1}{2\pi i}\oint_{\left| s\right| =R}\frac{1}{s}\left( \ln 
\frac{-s}{\mu ^{2}}\right) ^{j}ds \\
&=&\ln ^{j}\frac{R}{\mu ^{2}}-\frac{\pi ^{2}}{6}j(j-1)\ln ^{j-2}\frac{R}{\mu
^{2}}
\end{eqnarray*}
See also Ref.\cite{integ}

\section{Data integral}

The BESII data for the total $e^{+}e^{-}$ annihilation cross section is
represented in figure 1.

\begin{figure}[pt] 
\unitlength 1mm
\begin{picture}(135,60)
\put(20,0){ \epsfxsize=9cm \epsfbox{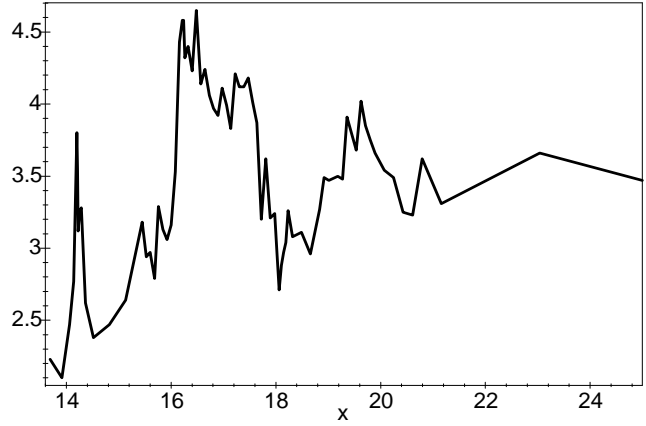}}
\end{picture}
\caption{$R(e^{+}e^{-}\rightarrow hadrons)$ from BES data as a function
  of $x=(cm-energy)^{2}$.} 
\label{Chfig1}
\end{figure}

The errors given by the authors of Ref. \cite{bessII} distinguish between
systematic and statistical errors. In our analysis this distinction is
absolutely essential as the statistical errors average out almost completely
in the integration process while the systematic errors prevail. The
threshold of the charm continuum is at twice the $D_{0}$ mass which
corresponds to $s_{0cont}=13.9\;GeV^{2}$. For definiteness we discuss here
only the integration up to $s_{\max }=25GeV^{2}$, the maximum value measured
by BESII. If we require for duality that the data within their errors agree
with QCD perturbation theory then it is seen that $s_{\max }$ may also be
chosen somewhat lower. We define

\begin{equation}
I_{1}=\int_{13.9}^{25}R_{continuum}(s)p(s)ds  \label{Idata}
\end{equation}

From this integral over the total cross section the contribution of the
light quarks must be subtracted to isolate the pure charm contribution. As $%
s>13.9GeV^{2}$ it is perfectly safe to use QCD perturbation theory for
contribution of the light flavors. We use the four loop result for the
massless correlator 
\begin{equation}
I_{2}=2\int_{13.9}^{25}\left( 1+\frac{\alpha _{s}}{\pi }+...\right) p\left(
x\right) dx  \label{Ilight}
\end{equation}
where terms up to $O(\alpha _{s}^{4})$ may be found in refs. \cite{Larin91}, 
\cite{Surg91},\cite{Chet97}. We use here the highest known order in $\alpha
_{s}$ because $I_{2}$ may be considered as an experimental input. The charm
continuum contribution is then given by 
\begin{equation}
I_{cont}=I_{1}-I_{2}  \label{Icont}
\end{equation}

Finally we have to take into account the two $J/\Psi $ and $\Psi ^{\prime }$
charmonium resonances below the continuum threshold 
\begin{equation}
I_{res}=9\pi (137.04)^{2}\left( m_{1}\Gamma _{1}p\left( m_{1}^{2}\right)
+m_{2}\Gamma _{2}p\left( m_{2}^{2}\right) \right)  \label{Ires}
\end{equation}
where the resonance masses and widths, are given by 
\begin{eqnarray}
m_{1} &=&3.0969GeV  \nonumber \\
\Gamma _{1} &=&(5.26\pm 0.37)\times 10^{-6}GeV  \label{pres} \\
m_{2} &=&3.6860GeV  \nonumber \\
\Gamma _{2} &=&(2.12\pm 0.18)\times 10^{-6}GeV.  \nonumber
\end{eqnarray}

The complete hadronic contribution to the sum rule is sum of these two
integrals 
\begin{equation}
I_{charm}=I_{cont}+I_{res}  \label{Ichar}
\end{equation}

\section{Polynomial weight functions}

There is much freedom in the choice of the polynomial in the duality
relation, Eq.(\ref{srule}). In this note we will use the polynomial to
reduce the importance of the continuum contribution relative to the well
established sub-threshold resonances. We impose the following conditions:

The polynomial $p(s)=1$ at $s=m_{J/\Psi }^{2}$, it vanishes at the end of
the integration range $(s=25Gev^{2})$, and it is small in the continuum
region. To be specific we choose an n-th order polynomial $%
p_{n}(s)=a_{0}+a_{1}s+..+a_{n}s^{n}$ and determine the coefficients by the
constraints 
\begin{eqnarray}
p_{n}(25) &=&0  \nonumber \\
p_{n}(m_{J/\Psi }^{2}) &=&1  \label{pcoef} \\
\int_{3.9}^{25}s^{k}p_{k}(s)ds &=&0, \quad k=0,1,..,n-1  \nonumber
\end{eqnarray}

\begin{figure}[htbp] 
\unitlength 1mm
\begin{picture}(135,60)
\put(25,0){ \epsfxsize=8cm \epsfbox{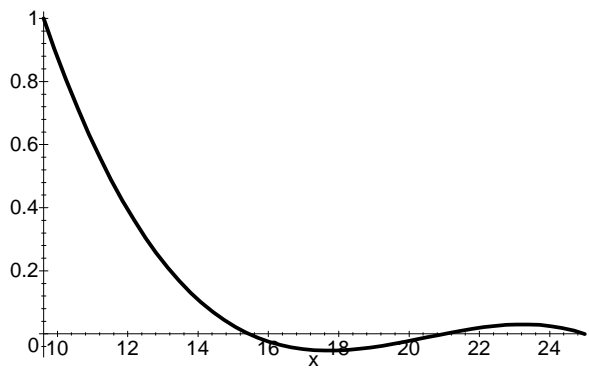}}
\end{picture}
\caption{Polynomial fit vanishing at the continuum range
  $(13.9-25)GeV^{2}$.} 
\label{fpol}
\end{figure}

For a 3-degree polynomial we obtain, for example 
\begin{eqnarray}
p_{3}(s) &=&7.\,\allowbreak 933\,787\,5-1.\,\allowbreak 209\,157\,911s 
\nonumber \\
&&+6.\,\allowbreak 015\,360\,076\times 10^{-2}s^{2}  \label{p3} \\
&&-9.\,\allowbreak 792\,537\,729\times 10^{-4}s^{3}  \nonumber
\end{eqnarray}
This result is plotted in Fig.\ref{fpol}.

It is obvious from the figure that the continuum data will almost cancel
when integrated with this polynomial.

\section{Results}

From the duality sum rule Eq.(\ref{srule}) we obtain with the help of Eqs.(%
\ref{Idata}-\ref{Ichar}) and Eqs.(\ref{Iqcd}, \ref{Icau}) 
\begin{equation}
I_{charm}=-\sum_{k=0}^{n}\sum_{i=0}^{6}\sum_{j=0}^{3}a_{k}A_{ij}(m_{c},\mu
)\left( m_{c}^{2}\right) ^{i}J(k-i,j)  \label{sreq}
\end{equation}
which can be solved for $m_{c}$.

We present here predictions for simple duality, i.e. $p(s)=1$, and for the
3rd-degree polynomial above. The unknown $m_{c}$ in Eq.(\ref{sreq}) is the
running charm mass at a scale $\mu $., which we fix to be $\mu =5GeV$, the
relevant scale of the problem.

For the coupling constant $\alpha _{s}$ we take as an input its value at the
mass of the tau lepton \cite{alpha} 
\begin{equation}
\alpha _{s}(m_{\tau })=0.345\pm 0.020  \label{3fcoup}
\end{equation}
with $m_{\tau }=1.777GeV.$ After appropriate matching \cite{Santa} from 3 to
4 flavors this corresponds to 
\begin{equation}
\alpha ^{4f}(5GeV)=0.224\pm 0.013  \label{4fcoup}
\end{equation}

\begin{figure}[htbp] 
\unitlength 1mm
\begin{picture}(135,60)
\put(25,0){ \epsfxsize=8cm \epsfbox{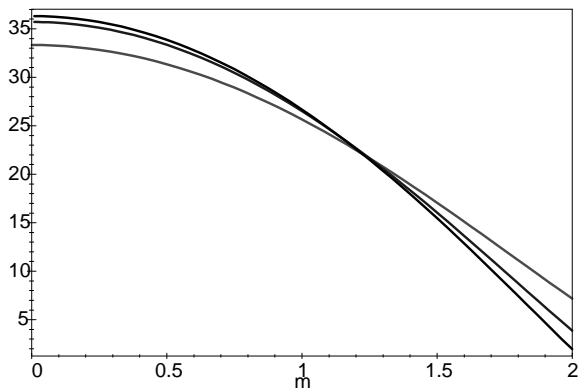}}
\end{picture}
\caption{QCD integral without polynomial weight as a function of
  mass. The three curves (from top to bottom at $m = 2$) represent tree,
  first and second order calculations in the strong coupling constant.}
\label{fQCD1}
\end{figure}

For simple duality, we plot in Fig.\ref{fQCD1} the contribution of the $QCD$
integral (lhs of Eq.(\ref{sreq})), at tree level and first and second order
in the strong coupling, as a function of the charm quark mass $m_{c}$.

For the Data integral we have the continuum contribution $%
I_{cont}=14.06GeV^{2}$ and the resonance contribution $I_{res}=12.80GeV^{2}$
with their sum being the total charm data integral 
\begin{equation}
I_{charm}=26.86GeV^{2}  \label{Ichar1}
\end{equation}

Solving Eq.(\ref{sreq}) for this value of charm data, the results for the
mass of the charm quark at different orders in the perturbative expansion
are: $m_{c}^{(0)}=0.916GeV$, $m_{c}^{(1)}=0.980GeV$ and $m_{c}^{(2)}=0.990GeV
$. We see that the convergence of the $QCD$ asymptotic expansion is
extremely good. The main source of uncertainties come from the strong
coupling constant and data. The result that we quote with this approach is 
\begin{equation}
m_{c}(\mu =5GeV)=(0.99\pm 0.01_{asymp}\pm 0.01_{\alpha _{s}}\pm
0.04_{res}\pm 0.10_{cont})GeV  \label{mc1}
\end{equation}
The asymptotic uncertainty comes from the difference of two- and three-loop
results used for the QCD correlators. The result of Eq. \ref{mc1}
corresponds to an invariant mass \cite{Larin97} 
\begin{equation}
m_{c}(m_{c})=(1.40\pm 0.11)GeV  \label{mm1}
\end{equation}
In this final result the errors have been added quadratically.

The contribution of the gluon condensate is completely negligible

\begin{figure}[htbp] 
\unitlength 1mm
\begin{picture}(135,60)
\put(25,0){ \epsfxsize=8cm \epsfbox{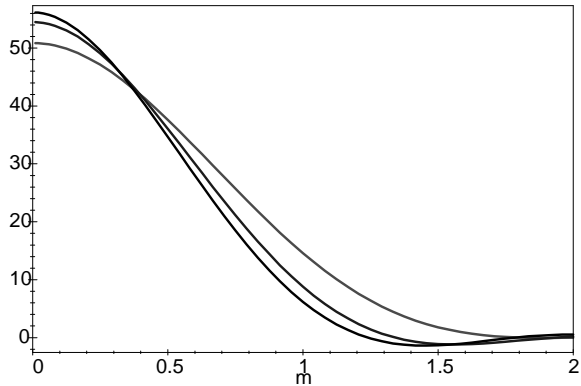}}
\end{picture}
\caption{QCD integral with polynomial weight as a function of mass. The
  three curves (from top to bottom at $m = 1$) represent tree, first and
  second order calculations in the strong coupling constant.}
\label{fQCD2}
\end{figure}

Our second approach consist in plug-in the 3rd degree polynomial in Eq.(\ref
{p3}) into Eq.(\ref{sreq}) in order to minimize the contribution from the
charm continuum data, and therefore minimize the error involved in these
data. As before we plot in Fig.\ref{fQCD2} the $QCD$ integral as a function
of $m_{c}$ for different orders in the perturbative expansion

The contribution from the continuum data is now $I_{cont}=-0.176Gev^{2}$,
whereas the resonance contribution, practically from $J/\Psi $ resonance, is 
$I_{res}=9.341GeV^{2}$. The complete charm data contribution is then 
\begin{equation}
I_{charm}=9.165GeV^{2}  \label{Icha2}
\end{equation}
With this value we solve again Eq.(\ref{sreq}) at different orders in the
perturbative expansion of $\alpha _{s}$ with the results: $%
m_{c}^{(0)}=1.153GeV$, $m_{c}^{(1)}=0.991GeV$, $m_{c}^{(2)}=0.927GeV$. We
see that, although we have eliminate the uncertainty coming from continuous
data, the price we pay is that the asymptotic expansion does not converge so
nicely as before, but still good enough to make a sensible prediction for
the mass of the charm quark. We have 
\begin{equation}
m_{c}(\mu =5GeV)=(0.93\pm 0.06_{asymp}\pm 0.015_{\alpha _{s}}\pm
0.014_{res})GeV  \label{mc2}
\end{equation}
This corresponds to an invariant mass \cite{Larin97} 
\begin{equation}
m_{c}(m_{c})=(1.34\pm 0.08)GeV  \label{mm2}
\end{equation}
The influence of the gluon condensate is in this approach $\sim 0.3\%$ for $%
\left\langle \alpha _{s}GG/\pi \right\rangle \sim .024GeV^{4}$, still
negligible.

The result we find is perfectly compatible, within error-bars, with the one
we found above by the simple duality approach. This agreement constitutes a
non trivial confirmation of the duality ansatz. Averaging the results of
both approaches, we finally  find

\begin{equation}
m_{c}(m_{c})=(1.37\pm 0.09)GeV.  \label{final}
\end{equation}

Our value for $m_{c}(m_{c})$ appears to be slightly bigger than the ones
given by alternative QCD sum rule methods. With our normalization point the
latter results read $m_{c}(m_{c})=(1.29\pm 0.05)GeV$\cite{NarisPL} and $%
m_{c}(m_{c})=(1.28\pm 0.06)GeV$\cite{Domin}. Although these values agree,
within error bars, with ours, it should be kept in mind that these authors
use older and lower values of the QCD coupling constant, so that the
error-bar quoted should really be larger. 

\section{Conclusions}

In this letter we have analyzed the mass of the charm quark in the context
of $QCD$ finite energy sum rules. In the phenomenological side of the sum
rule we use recent BESII $e^{+}e^{-}$ data, whereas in the theoretical side
we employ the a large momentum expansion of $QCD$ vector correlator function
up to $O(\alpha _{s}^{2})$ and $O(m_{c}^{12}/q^{12})$ . Two approaches are
considered. The first one uses simple Cauchy sum rule for the correlator.
The second one includes a polynomial in the sum rule to minimize the
contribution of the continuum data. The results from both approaches are
nicely compatible with each other. Whereas with the first approach suffers
from a substantial uncertainty arising from the continuum data, the second
one shifts this uncertainty to $QCD$ asymptotic expansion. The results allow
a nice consistency check of $QCD$ duality assumption. More precise results
need either better data or further terms in $QCD$ asymptotic
expansion.

\newpage

\begin{thebibliography}{99}
\bibitem{bessII}  J. Z. Bai et al. (BES collaboration), hep-ex/0102003, Feb
2001.

\bibitem{NdeR}  S. Narison and E. de Rafael. Nucl. Phys. B169 (1980) 253.

\bibitem{Chet1}  K. G. Chetyrkin, R. Harlander, J. H. K\"{u}hn and M.
Steinhauser, Nucl.Phys. B503 (1997) 339.

\bibitem{Chet2}  K. G. Chetyrkin, R. V. Harlander and J. H. K\"{u}hn,
Nucl.Phys. B586 (2000).

\bibitem{Jamin}  M. Eidemuller and M. Jamin, Phys.Lett. B498 (2001) 203.

\bibitem{Naris}  S. Narison, Nucl.Phys.Proc.Suppl. 74 (1999) 304.

\bibitem{NarisPL}  S. Narison, Phys.Lett. B341 (1994) 73.

\bibitem{Domin}  C. A. Dominguez, G. R. Gluckman and N.Paver, Phys.Lett.
B333 (1994) 184.

\bibitem{pineda} Antonio Pineda, ''Determination of the bottom quark
mass from the $\Upsilon(1S)$ system'', University of Karlsruhe preprint
Nr. TTP01-12, May 2001, hep-ph 0105008. 

\bibitem{Kuhn}  J. H. K\"{u}hn and M. Steinhauser, Phys.Lett. B437 (1998)
425.

\bibitem{ACD}  S. Groote, J. G. K\"{o}rner, K. Schilcher, N. F. Nasrallah,
Phys.Lett. B440 (1998) 375.

\bibitem{Gcond}  D. J. Broadhurst, P.A. Baikov, J. Fleischer, O. V. Tarasov
and V. A. Smirnov, Phys.Lett. B329 (1994) 103.

\bibitem{integ}  N. A. Papadopoulos, J. A. Pe\~{n}arrocha, F. Scheck and K.
Schilcher, Nucl.Phys. B258 (1985) 1.

\bibitem{Larin91}  S. G. Gorishny, A. L. Kataev and S. A. Larin, Phys. Lett.
259B (1991), 144.

\bibitem{Surg91}  L. R. Surguladze and M. A. Samuel, Phys.Rev.Lett. 66
(1991) 560; 66 (1991) 2416(E).

\bibitem{Chet97}  K. G. Chetyrkin, Phys. Lett. 391B (1997) 402.

\bibitem{alpha}  A. Pich, CERN Courier 40 (2000) 20.

\bibitem{Santa}  G. Rodrigo, A. Pich and A. Santamaria, Phys.Lett. B424
(1998) 367.

\bibitem{Larin97}  J. A. M. Vermaseren, S. A. Larin and T. van Ritbergen,
Phys.Lett. B405 (1997) 327.

\end{thebibliography}
\end{document}